\documentclass[12pt]{JHEP} 
\usepackage{epsfig} 
\newcommand{\Frac}[2]{\frac{\displaystyle #1}{\displaystyle #2}}      
\title{A phenomenological description of  $K\rightarrow
\pi\pi\gamma$ magnetic transitions \thanks{Work supported in part by TMR,
EC--Contract No. ERBFMRX-CT980169
(EURODA$\Phi$NE).}}

\author{Giancarlo D'Ambrosio\\Istituto
Nazionale di Fisica Nucleare,
Sezione di Napoli, Dipartamento di Scienze Fisiche, Universit\`a di Napoli
I-80126 Napoli, Italy\\ \email{E-mail: Giancarlo.Dambrosio@na.infn.it}}
\author{Dao-Neng Gao\thanks{On leave from {\it the Department of
Astronomy and Applied Physics, University of Science and Technology of 
China, Hefei, Anhui, 230026, China.}}\\ Istituto Nazionale di Fisica
Nucleare, Sezione di Napoli, Dipartamento di Scienze Fisiche, Universit\`a
di Napoli I-80126 Napoli, Italy\\ \email{E-mail: gao@na.infn.it}}

\received{}
\accepted{}
\abstract{
A phenomenological analysis of $K\rightarrow \pi\pi\gamma$
($K_L\rightarrow\pi^+\pi^-\gamma$ and $K^+\rightarrow\pi^+\pi^0\gamma$)
 with the direct emission photon is carried out  beyond the leading order
in the chiral perturbation theory. We show
that the experimental evidence for the large photon energy dependence of
the magnetic amplitude   
in $K_L\rightarrow\pi^+\pi^-\gamma$ seems to
indicate an interesting consequence: vector meson dominance must be
implemented at $O(p^4)$, which is not a general feature of the chiral
perturbation theory. The phenomenology of
$K^+\rightarrow\pi^+\pi^0\gamma$ is also analyzed using the same scheme.
}


\preprint{INFNNA-IV-2000/30\\October 2000}
\begin{document}

\section{Introduction}

Non-leptonic kaon decays have been an important tool for studying the
weak interactions \cite{GI98, Rafael, BBL96, ENP94, BK00}. 
The radiative non-leptonic kaon decays such as the processes
$K\rightarrow \pi\pi\gamma$ ($K_L\rightarrow\pi^+\pi^-\gamma$ 
and $K^+\rightarrow\pi^+\pi^0\gamma$) are
dominated by long distance contributions, and it is not easy to match
long and short distance contributions of these processes. 
In Ref. \cite{Rafael}, an impressive improvement for the evaluation of the
weak matrix elements in some particular processes matching long and short
distance contributions is obtained in the large $N_C$ limit. Spin-1
resonances are implemented in this context and important to achieve the
matching. Here we somewhat complement their work by looking for a good low
energy phenomenological description of the spin-1 resonances in the weak
sector to study the decays of $K\rightarrow\pi\pi\gamma$.

The total amplitude of $K\rightarrow \pi\pi\gamma$ contains two kinds of
contributions: the inner bremsstrahlung (IB) and direct
emission (DE).
Due to the pole in the photon energy the IB amplitude generally
dominates unless the non-radiative one is suppressed due to some
particular reason. This is the case of $K_L\rightarrow \pi^+\pi^-\gamma$
and $ K^+\rightarrow \pi^+ \pi^0\gamma$. The
non-radiative amplitude of
the former is suppressed by CP invariance and the latter one is suppressed
due to the $\Delta I=1/2 $ rule. It is of interest to extract the DE
amplitude of
these channels in order to reveal the chiral structure of the processes.  
DE contribution can be decomposed into
electric and magnetic parts in a multipole expansion \cite{LV88,
DMS92}. The
available
experimental evidence is consistent with a dominant magnetic part for the
DE amplitude \cite{Abrams72, Ram93}. So we will focus our attention on the
magnetic part.

In the framework of chiral perturbation theory($\chi$PT) \cite{KMW90, EKW93}, 
$K\rightarrow \pi\pi\gamma$ has been analyzed previously
\cite{ENP94,LV88, DMS92,DMS93,DI95,Cheng90,DEIN95,DP98},
and the leading order
of the magnetic amplitudes of the processes, starting at $O(p^4)$, 
appear as a constant with two kinds of
contributions: i) the reducible type from Wess-Zumino-Witten
action and $O(p^2)$ weak lagrangian ${\cal L}_2^{\Delta S=1}$; ii) the
local type from  $O(p^4)$ weak lagrangian ${\cal L}_4^{\Delta S=1}$. 
However, experimental analysis has
found a clear and large dependence on the photon energy in
$K_L\rightarrow
\pi^+\pi^-\gamma$ \cite{Ram93,KTeV}. In the $K^+$ case, the energy
dependence of the
magnetic amplitude has not been observed yet. In order to
explain the photon energy dependence in $\chi$PT, the theoretical
analysis has to be beyond the leading order \cite{ENP94,DEIN95,DP98}.  
A complete $O(p^6)$ magnetic amplitude of         
$K\rightarrow\pi\pi\gamma$ generally may be useful in the future but not  
now since some  unknown parameters have to be introduced, which in fact   
makes the prediction impossible. 

The recent direct measurement of the
$K_L\rightarrow\pi^+\pi^-\gamma$ DE form-factor by KTeV Collaboration
\cite{KTeV} clearly indicates a vector meson dominance (VMD) form-factor
\begin{equation}
{\cal F}=\Frac{A_1}{1-\frac{m_K^2}{m_V^2}+\frac{2m_K}{m_V^2} 
{E_\gamma^*}}+A_2,
\label{EVMD}
\end{equation}
where $A_1$ and $A_2$ are constants with $A_1/A_2=-1.243\pm 0.057$
, and $E_\gamma^*$ is the photon energy in
the $K_L$ rest frame. 
Eq. (\ref{EVMD}) gives the best $\chi^2$ for a single-parameter fit
 ($\chi^2$/DOF is 38.8/27),
compared with the linear slope fit ($\chi^2$/DOF is 43.2/27), and
two-parameter quadratic slopes fit ($\chi^2$/DOF is 37.6/26).
Therefore, this measurement seems to indicate that VMD
should be implemented at $O(p^4)$ instead of being at $O(p^6)$ for this
decay.  This point is not well understood currently within $\chi$PT. 
The purpose of the present paper is to understand it  
using a phenomenological description. Also, we extend our analysis
to the decay $K^+\rightarrow\pi^+\pi^0\gamma$.

In Section 2, we remind briefly the kinematics of
$K\rightarrow \pi\pi \gamma$.  In Section 3, we carry out 
a phenomenological analysis of $K\rightarrow \pi\pi \gamma$ 
using chiral lagrangian plus VMD. 
The results are summarized in Section 4. 

\section{Kinematics}

The general invariant amplitude of $K\rightarrow \pi\pi\gamma$ can be
defined as follows \cite{ENP94,DI95}
\begin{equation}
A[K(p)\rightarrow\pi_1(p_1)\pi_2(p_2)\gamma(q,\epsilon)]=\epsilon^\mu(q)
M_\mu (q,p_1,p_2), \end{equation} 
where $\epsilon_\mu(q)$ is the photon polarization and $M_\mu$ is
decomposed into an electric $E$ and a magnetic $M$ amplitudes as    
\begin{equation} 
M_\mu={\frac{E(z_i)}{m_K^3}}[p_1{\cdot}q p_{2\mu}-p_2{\cdot}q
p_{1\mu}]+{\frac{M(z_i)}{m_K^3}} \epsilon_{\mu\nu\alpha\beta} p_1^\nu
p_2^\alpha q^\beta,
\end{equation}  
with 
\begin{eqnarray*}
z_i=\frac{q{\cdot}p_i}{m_K^2},\;(i=1,2),\;\; z_3=\frac{p{\cdot}q}{m_K^2}
,\;\;z_3=z_1+z_2. 
\end{eqnarray*} 
The double differential rate for the unpolarized photon is
\begin{equation}
\frac{\partial^2\Gamma}{\partial z_1\partial z_2}=\frac{m_k}{(4 \pi)^3}
(|E(z_i)|^2+ |M(z_i)|^2) [z_1 z_2(1-2 z_3-r_1^2-r_2^2)-r_1^2 z_2^2-r_2^2
z_1^2],
\end{equation} 
where $r_i=m_{\pi_i}/m_K$. 

In $K_L\rightarrow \pi^+\pi^-\gamma$, the most useful variables  are: (i) the photon
energy in the kaon rest frame $E_\gamma^*$, and  (ii) the angle $\theta$ between the
photon and $\pi^+$ momenta in the di-pion rest frame. The relations between
$E_\gamma^*$, $\theta$ and the $z_i$ are:    
\begin{eqnarray}  
z_3=\frac{E_\gamma^*}{m_K},\;\;\;\;z_{\pm}=\frac{E_\gamma^*}{2m_K}(1\mp\beta {\rm cos}
\theta),
\end{eqnarray}  
where $\beta=\sqrt{1-4m_\pi^2/(m_K^2-2m_K E_\gamma^*)}$.
Then the differential rate is 
\begin{eqnarray} 
\frac{\partial^2 \Gamma}{\partial E_\gamma^*\partial   
{\rm  cos}\theta}=\frac{(E_\gamma^*)^3\beta^3}
{512\pi^3m_K^3}\left(1-\frac{2E_\gamma^*}{m_K}\right){\rm sin}^2\theta(|E|^2+|M|^2). 
\end{eqnarray} 

For $K^+\rightarrow\pi^+\pi^0\gamma$, three photons will be detected in
the measurement, so it is more useful to study
the differential rate as a function of: (i) the charged pion kinetic
energy in  
the $K^+$ rest frame $T_c^*$, and (ii) $W^{2}=(q{{\cdot} }p_{K})(q\cdot p_{+})/(m_{\pi
^{+}}^{2}m_{K}^{2})$ \cite{Abrams72}. These two variables are related to
the $z_{i}$ by 
\begin{eqnarray}
&&z_{0}=\frac{1}{2m_{K}^{2}}(m_{K}^{2}+m_{\pi    
^{+}}^{2}-m_{\pi^{0}}^{2}-2m_{K}m_{\pi ^{+}}-2m_{K}T_{c}^{\ast }),     
\label{Z0TC}
\\
&&z_{3}z_{+}=\frac{m_{\pi ^{+}}^{2}}{m_{K}^{2}}W^{2}.      
\end{eqnarray}
The advantage of using these variables is that, through the $W^{2}$  
dependence, one can easily disentangle the different contributions of  
the IB, DE
amplitudes, and interference term between IB and DE  
\begin{eqnarray}
\Frac{\partial ^{2}\Gamma }{\partial T_{c}^{\ast }\partial W^{2}}   
&=&\Frac{\partial ^{2}\Gamma _{IB}}{\partial T_{c}^{\ast }\partial
W^{2}}\left[1+\Frac{m_{\pi ^{+}}^{2}}{m_{K}}2Re\left(\Frac{E_{DE}}{eA}\right)W^{2}   
\right. \nonumber \\   
&&\left. +\Frac{m_{\pi^{+}}^{4}}{m_{K}^{2}}\left(\left|\Frac{E_{DE}}{eA}\right|^{2}+\left|\Frac{M_{DE}}{    
eA}\right|^{2}\right)W^{4}\right],
\end{eqnarray}
where $A=A(K^{+}\rightarrow \pi ^{+}\pi ^{0})$. 

\section{Analysis}

The leading order magnetic amplitudes of $K\rightarrow\pi\pi\gamma$
start at $O(p^4)$
in $\chi$PT\cite{ENP94,DEIN95}
\begin{eqnarray}
M^{(4)}_{L}=\frac{e G_8 m_K^3}{2\pi^2 F}(a_2+2a_4),\label{ML4}\\
M^{(4)}_{+}=-\frac{e G_8 m_K^3}{4\pi^2 F}[2+3(2 a_3-a_2)].\label{MP4}
\end{eqnarray}
The subscripts $L$ and $+$ denote  $K_L\rightarrow\pi^+\pi^-\gamma$ and
$K^+\rightarrow\pi^+\pi^0\gamma$ respectively. The $a_i$'s parts of the
above amplitudes come from the local weak lagrangian ${\cal
L}_4^{\Delta S=1}$. The first term in $M^{(4)}_{+}$ is the reducible type
contribution.  Due to the Gell-Mann-Okubo mass relation, the reducible
magnetic amplitude of $K_L\rightarrow\pi^+\pi^-\gamma$ generated from
$K_L-\pi^0 (\eta_8)$ mixing vanishes at $O(p^4)$. However, when
$\eta^\prime$ is included, thus $\eta-\eta^\prime$ mixing is considered,
there is a reducible amplitude called $F_1$ term in Refs. \cite{ENP94,
DP98}, which is therefore at $O(p^6)$. 

The experimental analysis of
$K_L\rightarrow\pi^+\pi^-\gamma$ using the VMD form-factor
parameterization by KTeV Collaboration \cite{KTeV} indicates that VMD must
be implemented at $O(p^4)$. This means that the couplings $a_i$'s, $i$=1,
2, 3, 4 (or in terms of $N_i$'s, $i$=28, 29, 30, 31) should get the
contribution from the vector resonance exchange.

\FIGURE{\epsfig{file=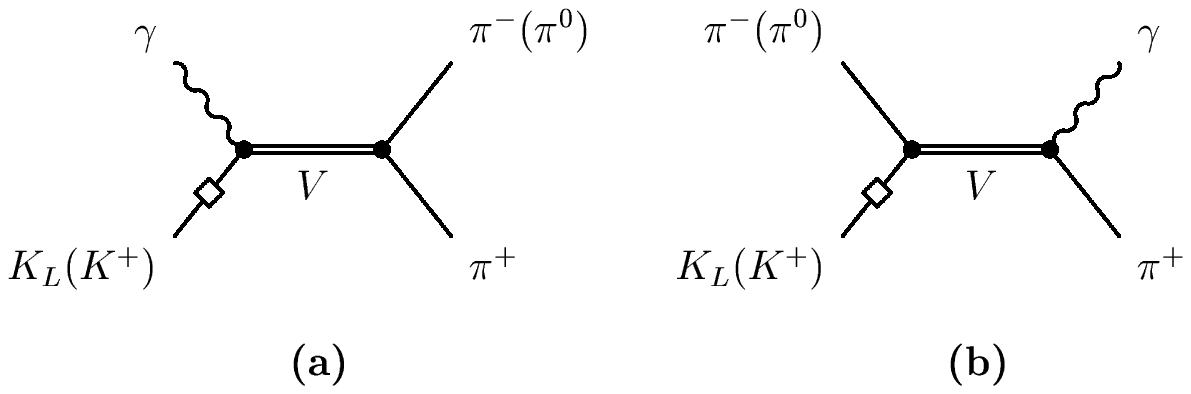,height=1.8in}
\caption{Diagrams contributing to the 
indirect VMD magnetic amplitude of $K_L\rightarrow\pi^+\pi^-\gamma$
or $K^+\rightarrow\pi^+\pi^0\gamma$. The
diamond in the external legs denotes $K_L-\pi^0,\eta_8$ or
$K^+-\pi^+$ mixing. The
black circle denotes the strong/electromagnetic vertex. The crossed
diagram $\pi^+\leftrightarrow\pi^-(\pi^0)$ should be considered in (b).}}

\FIGURE{\epsfig{file=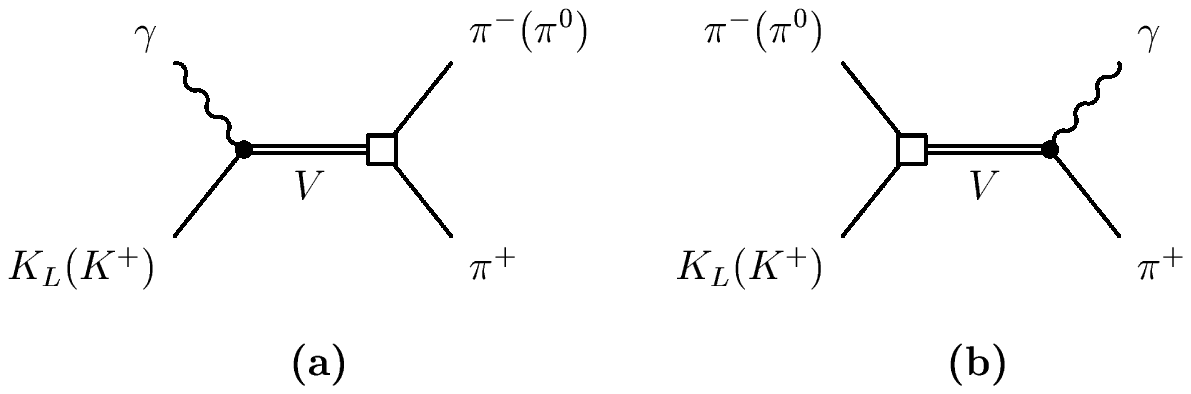,height=1.8in}
\caption{Diagrams contributing to the direct VMD magnetic
amplitude of $K_L\rightarrow\pi^+\pi^-\gamma$ or
$K^+\rightarrow\pi^+\pi^0\gamma$. The empty box denotes
the vertex
generated by eq. (\ref{VPP}), and the black circle denotes the
strong/electromagnetic vertex. The crossed diagram
$\pi^+\leftrightarrow\pi^-(\pi^0)$ should be considered in (b).}}

VMD can be introduced into the effective lagrangian phenomenologically,
and there are two different kinds of vector resonance exchange
contributions to the non-leptonic radiative kaon decays \cite{EPR90, DP97}:

(i) Vector resonance exchange between strong/electromagnetic vertices with
a weak transition in an external leg, as shown in Fig. 1, which are
usually called as indirect
transitions. The amplitude from this kind of transition is of reducible
type, and  vanishes at $O(p^4)$, which therefore contributes first at
$O(p^6)$.  

(ii) The direct weak transitions are those where the weak vertices
involving the vector resonances are present, which could 
contribute to the couplings $a_i$'s.
But this is NOT a general feature of VMD.
Indeed the antisymmetric tensor realization of vector resonances does not
generate any vector exchange contributions to $K\rightarrow \pi\pi\gamma$
at $O(p^4)$. However, this is not the case for the other realization
approaches such as massive Yang-Mills, hidden local symmetry, and
conventional vector formulations.  It has been pointed out in
Ref. \cite{EPR90} that, for the odd-intrinsic parity operator relevant in
the $V\rightarrow P\gamma$, the antisymmetric tensor formulation would
give contributions starting at $O(p^4)$ while QCD requires an explicit
$O(p^3)$ term given by the conventional vector formulation.  
As already realized in
\cite{DP98A}, using the conventional vector formulation, there are
$O(p^4)$ VMD
contributions generated by the following
operators through the direct weak transition (see Fig. 2):
\begin{equation}
{\cal L}_R^{O(p)}=G_8 F^4[\omega_1^R\langle \Delta
\{R_\mu,u^\mu\}\rangle+\omega^R_2\langle\Delta u_\mu\rangle \langle
R^\mu\rangle],
\label{VPP}
\end{equation}
where $R=V,A$, denoting the vector and axial-vector resonances
respectively. In the factorization, the couplings $\omega_i^R$ are
\begin{equation}
\omega_1^R=-\omega_2^R=\sqrt{2}\frac{m_R^2}{F^2}f_R \eta_R,
\label{factorization}
\end{equation}
with $\eta_R$ is the factorization parameter. 
$f_V$ and $f_A$  are the effective
couplings in the general strong/electromagnetic lagrangian involving
spin-1 resonances (We use the notations in Ref. \cite{DP98}).
As shown in Refs. \cite{DP98A, DP98}, the spin-1 resonance
contributions to $a_i$'s have been obtained.
Indeed, the operators in eq. (\ref{VPP}) with $R=V$ do generate the
structure of the VMD form-factor starting at $O(p^4)$ once the full
propagator of the vector resonance is taken into account (the axial-vector
only contributes a constant). The use of the full VMD propagator was
also suggested in $K\rightarrow\pi\gamma^*$ \cite{DEIP}.

The next leading order magnetic amplitude of $K\rightarrow\pi\pi\gamma$
in $\chi$PT is at $O(p^6)$, which contains
two parts: local contribution and loop contribution. Although the general 
local $O(p^6)$ couplings in weak effective lagrangian have not been
developed yet, we can be sure that many  unknown parameters have to
appear, which will make the prediction  impossible. One may expect that
the local terms could be generated through resonance exchange which are
reasonably thought as the most relevant ones, and the large number of
unknown couplings could be reduced significantly.  But a complete
determination of the contributions from resonances including vector,
axial-vector, scalar, and pseudoscalar still remains very difficult
because we have to face some unknown pseudoscalar-resonance weak couplings
which cannot be fixed by experiments.  Also, at $O(p^4)$, we know that 
the couplings $a_i$'s have other contributions than that from the
resonances: 
for instance, there   
exists the contribution from the WZW anomaly action, giving $0<a_i^{\rm an}
\le 1.0$ \cite{Cheng90, BEP92} with $a_i^{\rm an}$ is 
the unknown parameter. In the $K_L$ case, there is $O(p^6)$ $F_1$
term, which is sensitive to the octet symmetry
breaking \cite{ENP94, DP98}.

On the other hand, one can reasonably assume that the photon 
energy dependence of the 
amplitude is dominated by the vector resonance exchange while the other
resonances including axial-vector, scalar and pseudoscalar only generate
the constants contributing to the amplitude. 
Moreover, we have checked the $O(p^6)$ loop would lead to very negligible
energy
dependent contribution in $K_L\rightarrow\pi^+\pi^-\gamma$ and $K^+\rightarrow\pi^+\pi^0\gamma$.  
Therefore, a phenomenological description is that, one can express 
the full magnetic amplitude as non-VMD part (which is the constant but with 
large uncertainties involved in it) 
and VMD part (which is energy dependent and could be determined up to
one parameter), and use the corresponding 
experimental decay rate to determine the former part. Here we
use the recent observed values of $K_L\rightarrow\pi^+\pi^-\gamma$ and
$K^+\rightarrow \pi^+\pi^0\gamma$:
\begin{eqnarray}
Br(K_L\rightarrow\pi^+\pi^-\gamma;E_\gamma^*>20{\rm MeV})_{\rm
DE}=(3.10\pm0.05)\times
10^{-5}~ \cite{KTeV}, \label{KLBr}
\end{eqnarray}       
and 
\begin{eqnarray}
Br(K^{+}\rightarrow \pi ^{+}\pi ^{0}\gamma;55{\rm MeV}\leq T_{c}^{\ast }\leq 90{\rm
MeV})_{\rm DE} =(4.7{{\pm} }0.8){{\times} }10^{-6} \cite{BNL00}.\label{KPBr}
\end{eqnarray}

The VMD part magnetic amplitude of $K_L\rightarrow \pi^+\pi^-\gamma$,
corresponding to Figs. 1 and 2,  gives 
\begin{eqnarray}
M^L_{\rm VMD}=\frac{e G_8 m_K^3}{2\pi^2
F}\tilde{r}\left(\Frac{\eta_V+\frac{m_K^2}{m_V^2}(1-2
z_3)}{1-\frac{m_K^2}{m_V^2}+\frac{2m_K^2}{m_V^2}z_3}+
\Frac{\frac{\eta_V}{2}-\frac{m_K^2}{m_V^2}z_3}{1-
\frac{m_K^2}{m_V^2}z_3}\right), 
\label{KLVMD}
\end{eqnarray}
with
\begin{eqnarray}
\tilde{r}=\frac{32\sqrt{2}\pi^2 f_V h_V}{3}, 
\end{eqnarray} 
where $h_V$ is the coupling in the general strong/electromagnetic
lagrangian involving spin-1 resonances \cite{DP98}; the $\eta_V$ part is
$O(p^4)$, and the rest is $O(p^6)$. 

We would like to give some remarks here:

(1)~The VMD form factor like eq. (\ref{EVMD}) was firstly suggested by 
Lin and Valencia
\cite{LV88}, and it is phenomenologically successful. However,
as already noted by
Picciotto \cite{Pic92}, theoretically, there exists some inconsistency 
in that version 
because the indirect VMD form factor should vanish at $O(p^4)$ but the $O(p^6)$
indirect vector exchange contribution is important in understanding the magnetic
transition of $K_L\rightarrow\pi^+\pi^-\gamma$ \cite{Pic92, ENP94}. Here, we
have included both direct and indirect VMD contributions to the form factor
eq. (\ref{KLVMD}) with the former one starting at $O(p^4)$ 
 and the latter one starting
at $O(p^6)$, which properly satisfies all the theoretical constraints. 

(2)~The second term in eq. (\ref{KLVMD}), divided by
$(1-\frac{m_K^2}{m_V^2} z_3)$, is absent in eq. (\ref{EVMD}). 
However, theoretically, it is generated by 
Fig. 1b (corresponding to $z_3$ part) and
Fig. 2b (corresponding to $\eta_V$ part), we
have no reason to  exclude it. In fact it constructively enhances the
slope of $z_3$, thus affects the rate and the spectrum
significantly. Note 
that, as a good approximation, we have used $z_+= z_-\simeq z_3/2$ in deriving this
term.

(3)~The $\omega^V_2$ term in eq. (\ref{VPP}) does not contribute to
$K_L\rightarrow\pi^+\pi^-\gamma$, so we do not need the factorization relation
eq. (\ref{factorization}) in deriving eq. (\ref{KLVMD}). This means the present
calculation in $K_L$ case is independent of the factorization. It is an almost
model-independent prediction.

The non-VMD part can be written as
\begin{eqnarray}
M^L_{\rm non-VMD}=\frac{e G_8 m_K^3}{2\pi^2 F} A^L, 
\end{eqnarray}
with
\begin{eqnarray}
 A^L=(a_2+2a_4)_{\rm non-VMD}+~{\rm other~~ contributions}.   
\end{eqnarray}

\TABLE{
\begin{tabular}{c c c c c c c c c c c}
\hline\hline\\
$\eta_V$&0.1&0.2&0.3&0.4&0.5&0.6&0.7&0.8&0.9&1.0 \\ \\ \hline \\
$A^L$&0.55&0.36&0.17&-0.01&-0.20&-0.38&-0.57&-0.75&-0.94&-1.13\\ \\
\hline\\
$A^+$&1.53&1.34&1.16&0.97&0.79&0.60&0.42&0.23&0.05&-0.13\\ \\  
\hline\hline
\end{tabular}
\caption{The quantities $A^L$ and $A^+$ extracted from the observed
branching ratio
eqs. (\ref{KLBr}) and (\ref{KPBr}) for the different $\eta_V$.}
}

The values of $A^L$ are displayed in the second line of Table 1 in the
range of $0.1\le
\eta_V\leq 1.0$. We find that the spectrum of the photon energy is not
very sensitive to the value of $\eta_V$, as shown in Fig. 3, and  they are 
in agreement with the one
generated from eq. (\ref{EVMD}), which gives the best $\chi^2$ fit to the
data in Ref. \cite{KTeV}.   From Fig. 3, the best value of $\eta_V$
is about 0.5, which is reasonably consistent with the preferred
$\eta_V\simeq0.3$ obtained in the factorization \cite{DP97}. The
difference is that, some other $O(p^6)$
contributions parametrized using $\eta_{\rm VP\gamma}$ and $\eta_{\rm
VPP}$ are considered there \cite{DP97, DP98}. Therefore, it seems that the
high order contributions could enhance the value of $\eta_V$.

\FIGURE{
\epsfig{file=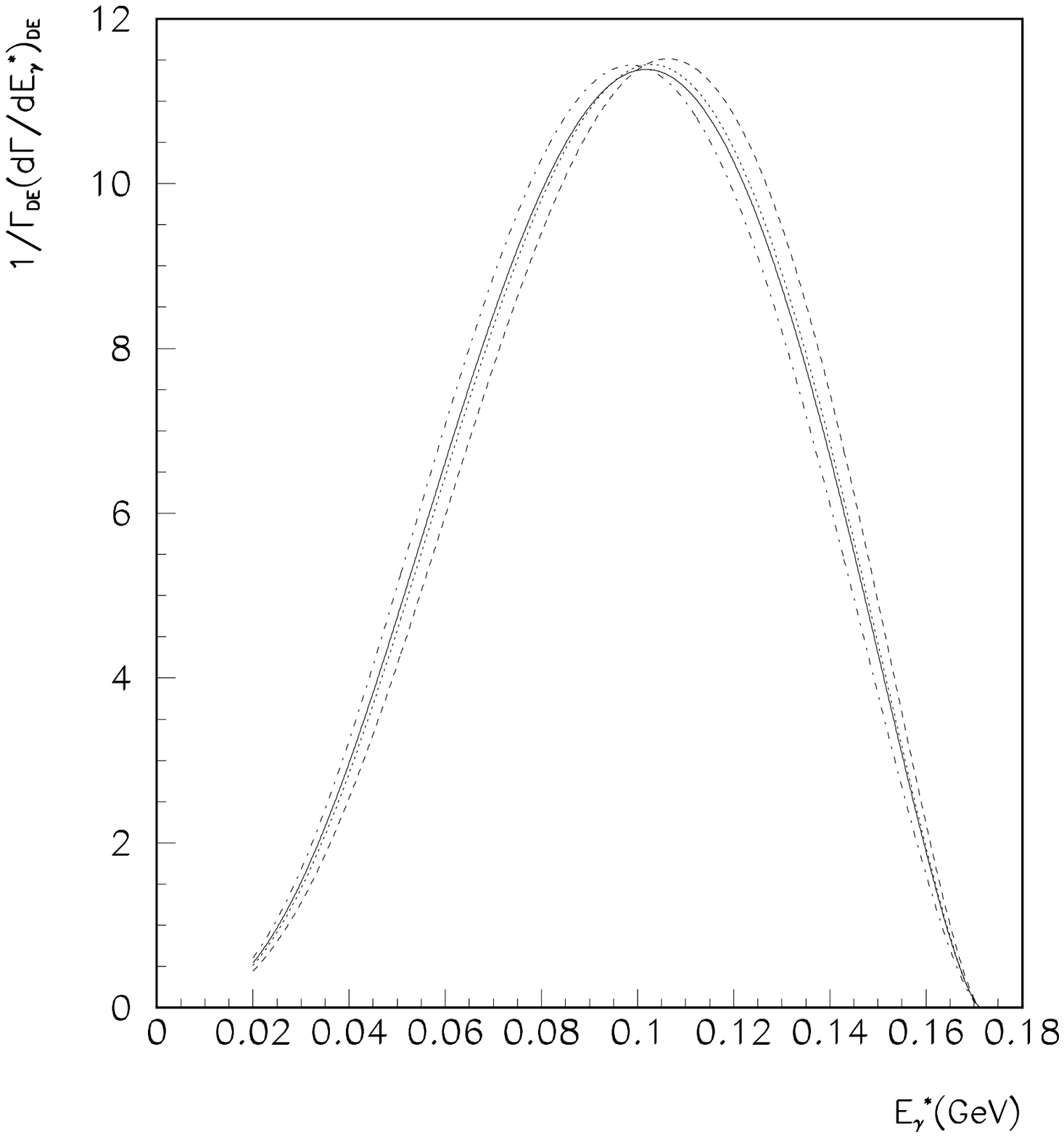,height=10cm,width=12cm}
\caption{$E_\gamma^*$ spectrum of the DE  
$K_L\rightarrow\pi^+\pi^-\gamma$ ($E_\gamma^*>20{\rm MeV}$). The
solid line is from eq. (\ref{EVMD}). The rest are generated from
eq. (\ref{KLVMD}): the dashed line corresponds to $\eta_V=0.1$, the dotted 
line corresponds to $\eta_V=0.5$, and the dot-dashed line corresponds to
$\eta_V=1.0$.} }

Likewise, in $K^+\rightarrow\pi^+\pi^0\gamma$, the VMD (from Figs. 1
and 2) and non-VMD parts
magnetic amplitudes  are
\begin{eqnarray}
M^+_{\rm VMD}=-\frac{e G_8 m_K^3}{4\pi^2
F}\tilde{r}\left (\Frac{\eta_V+\frac{m_K^2}{m_V^2}(1-2z_3)
}{1-\frac{m_K^2}{m_V^2}+\frac{2m_K^2}{m_V^2}z_3}+
\Frac{-\frac{\eta_V}{2}+\frac{2m_K^2}{m_V^2}z_+}{1-\frac{2m_K^2}{m_V^2}z_+}
+\Frac{\eta_V+\frac{2m_K^2}{m_V^2}z_0}{1-\frac{2m_K^2}{m_V^2}z_0}\right), 
\label{KPVMD}\nonumber \\
\end{eqnarray}
and 
\begin{equation}
M^+_{\rm non-VMD}=-\frac{e G_8 m_K^3}{4\pi^2  F}A^+,
\end{equation}
with
\begin{equation}
A^+=2+3(2a_3-a_2)_{\rm non-VMD}+~{\rm other~~ contributions}. 
\end{equation}

We have shown in Table 1 the values of $A^+$ for the different $\eta_V$,
and plotted the $T_c^*$ normalized to 
$m_K$ and $W$ spectrum
from the DE magnetic amplitude in Figs. 4 and 5, and the $W$ spectrum from
the sum of IB
and DE amplitude normalized to the IB spectrum in Fig. 6.
 Also,we find that these spectra are not sensitive to the $\eta_V$,
and the last one can be compared with the corresponding experimental
result \cite{BNL00}.   

\FIGURE{
\epsfig{file=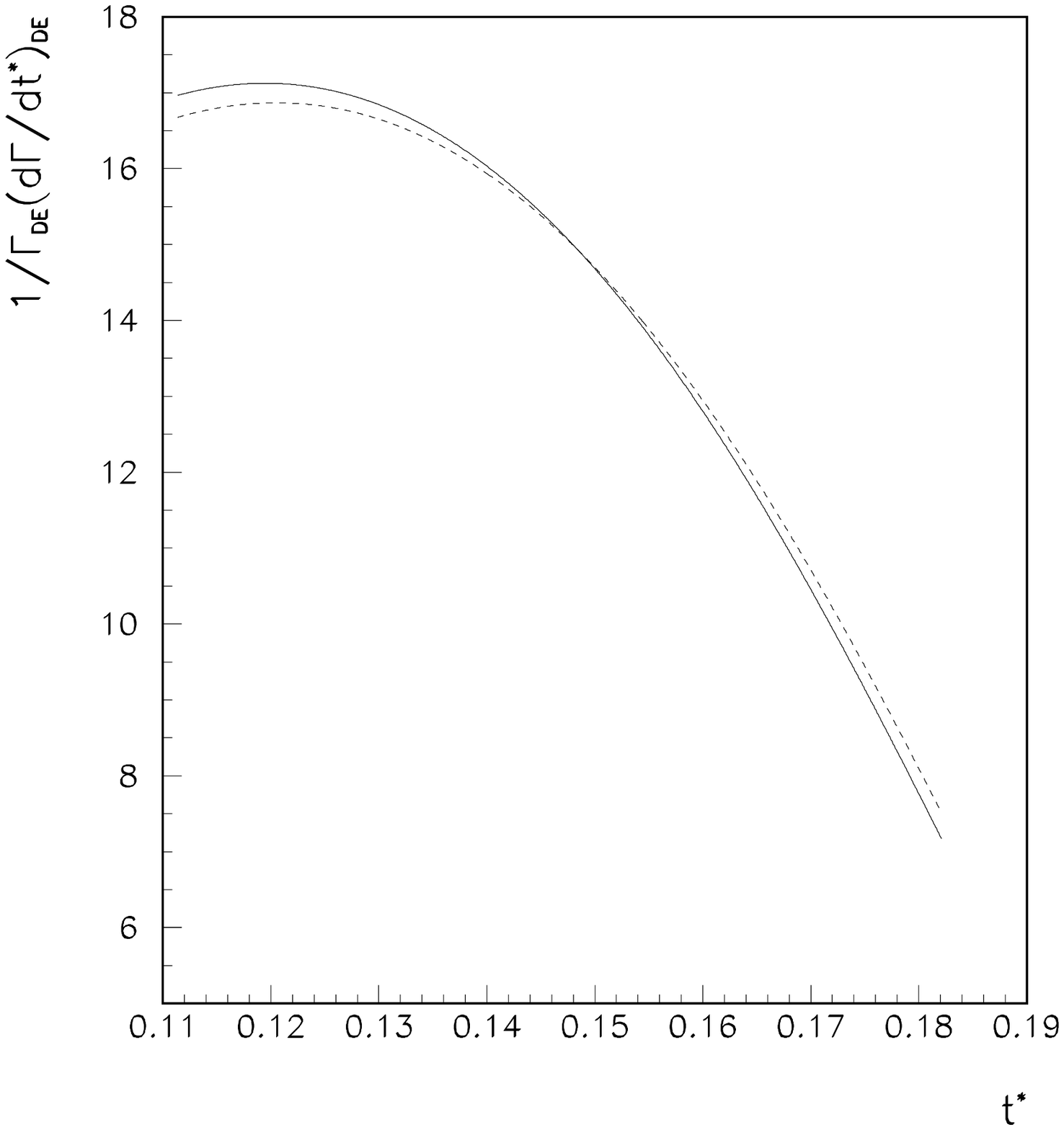,height=10cm,width=12cm}
\caption{Spectrum in $t^*$(=$T_c^*/m_K$) of the DE
$K^+\rightarrow\pi^+\pi^0\gamma$ with 55 MeV$\le T_c^*\le$90 MeV. The
solid line corresponds to
$\eta_V=0.1$. The dashed line corresponds to $\eta_V=1.0$.}
}

\FIGURE{
\epsfig{file=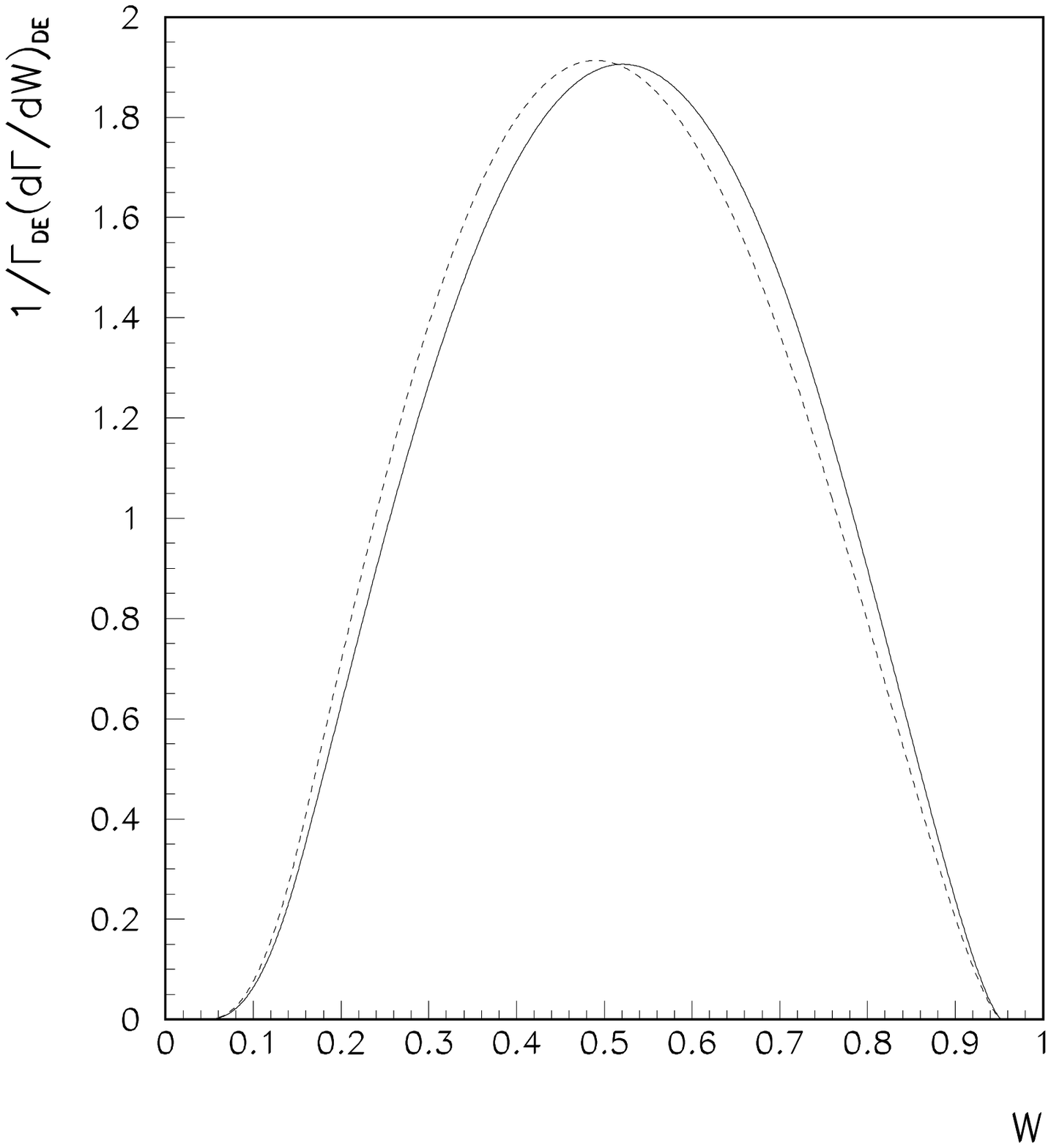,height=9cm,width=11cm} 
\caption{
Spectrum in $W$ of the DE $K^+\rightarrow\pi^+\pi^0\gamma$. The solid line
corresponds to $\eta_V=0.1$. The dashed line corresponds to
$\eta_V=1.0$.} 
}

\FIGURE{
\epsfig{file=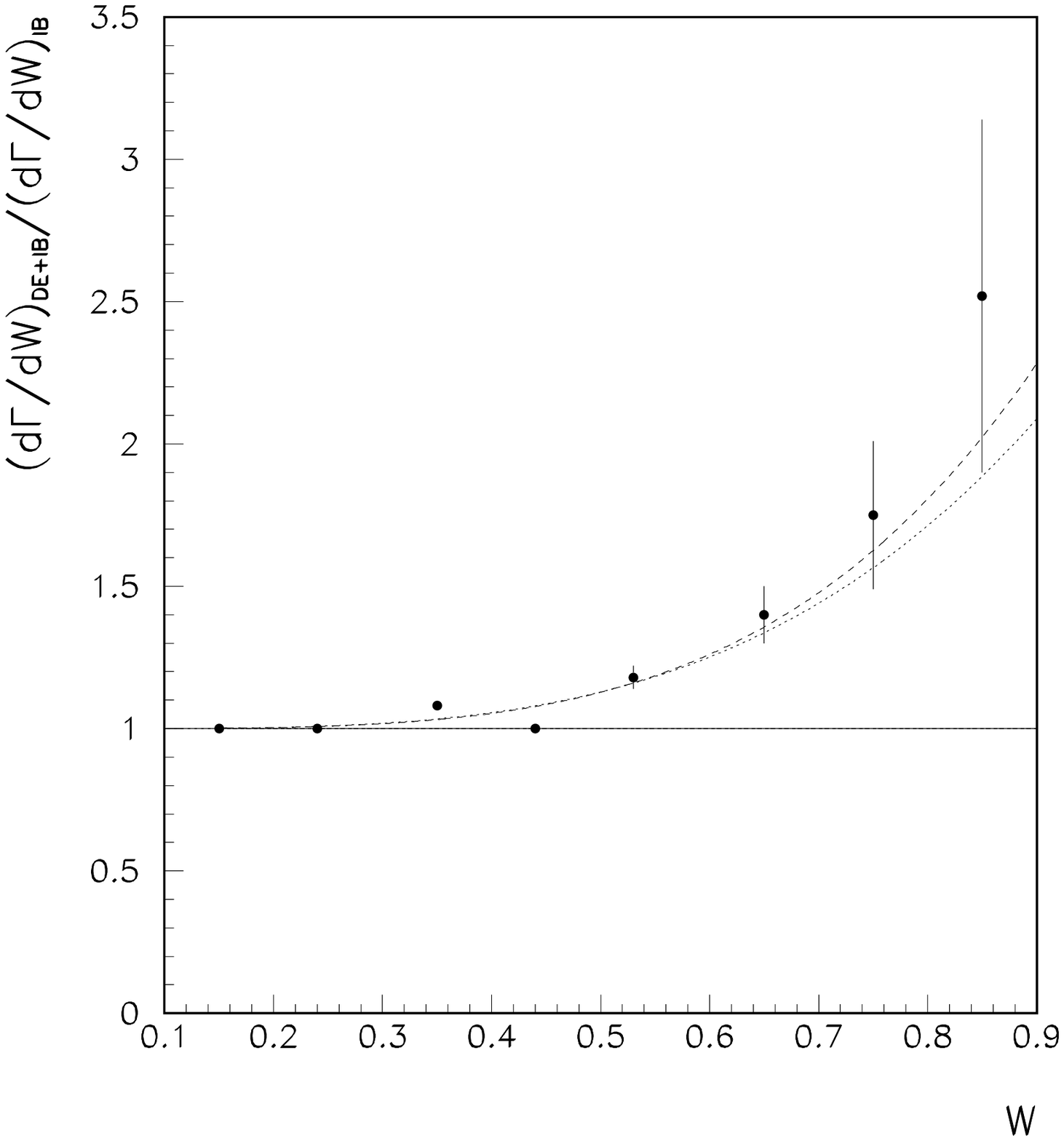,height=9cm,width=11cm} 
\caption{$W$ spectrum normalized to the IB spectrum of
$K^+\rightarrow\pi^+\pi^0\gamma$.  The dashed line corresponds to
$\eta_V=0.1$. The dotted line corresponds to $\eta_V=1.0$. The data points
are from Ref. \cite{BNL00}.}
}

\section{Conclusions}

We have presented a phenomenological description of the magnetic
amplitudes of $K_L\rightarrow\pi^+\pi^-\gamma$ and
$K^+\rightarrow\pi^+\pi^0\gamma$  beyond the leading order in $\chi$PT.
The VMD contribution plays an important role in the analysis. 
We parameterize the VMD part magnetic amplitudes of these two decays in
eqs. (\ref{KLVMD}) and (\ref{KPVMD}), and the non-VMD parts of the
amplitudes are estimated by fitting the corresponding observed decay
rates.  Our phenomenological description is consistent with the
factorization prediction $\eta_V\simeq 0.3$. 

We summarize the analysis as follows.

(1)~ We get the values of $A^L$ and $A^+$ in the range of $0.1\le\eta_V\le
1.0$ (see Table 1). $A^L$ is equals to $(a_2+2
a_4)_{\rm non-VMD}$
plus other higher order contributions. We know some other high order
contributions, for instance,  $F_1$ term is important and very sensitive
to the octet symmetry breaking \cite{ENP94, DP98}. So here we cannot
expect the conclusive information on $a_2+2 a_4$ from $A^L$. The
situation in $K^+$
case seems a little better. After neglecting higher order contributions to
$A^+$, we can get $-0.71\le (2a_3-a_2)_{\rm non-VMD}\le -0.1$.  From
Ref. \cite{DP98A} in the factorization, $(2a_3-a_2)_{\rm
axial-vector}\simeq 0.3 \eta_A$ with $0<\eta_A\le 1.0$ is the
factorization parameter.  If we assume the rest contribution 
to $(2a_3-a_2)$ is dominated by the one from WZW anomaly action, we find
our prediction on $(2a_3-a_2)^{\rm an}$ is consistent with the
expected $0<a_i^{\rm
an}\le 1.0$.

(2)~Although from our analysis $A^L$ could be positive or negative, the
large CP asymmtry $B_{CP}$ \cite{EWS95} in $K_L\rightarrow\pi^+\pi^-
e^+e^-$ 
originated from the interference between the magnetic and IB
amplitude of $K_L\rightarrow \pi^+\pi^-\gamma^*$ is predicted to be always
positive and not sensitive to $\eta_V$ in the present analysis,
which is consistent with the measurement \cite{KTeV98}. This is
not surprising if we carry out Taylor expansion over the form-factor in
eq. (\ref{KLVMD}) (we assume we can do this expansion), and express 
the total amplitude as
\begin{equation}
M=\frac{e G_8 m_K^3}{2\pi^2 F} \tilde{m}(1+r z_3+s z_3^2),     
\end{equation}  
we will get, in the range of $0.1\le\eta_V\le 1.0,$
\begin{eqnarray}
1.47\le \tilde{m} \le 1.75,\nonumber\\
2.08\le -r\le 2.88,\nonumber\\
2.50\le s \le 3.93,
\end{eqnarray}
which are comparable with the recent KTeV measurement 
$r=-2.93\pm0.41\pm0.34$, $s=3.31\pm1.15\pm 0.96$ \cite{KTeV}, and
$|\tilde{m}|=1.53\pm0.25$ in Ref. \cite{Ram93} from only linear slope fit. 

(3)~So far, there is no experimental evidence for the energy dependence
of the magnetic amplitude in $K^+\rightarrow
\pi^+\pi^0\gamma$. But the VMD form-factor obviously indicates this
energy dependence. By Taylor expansion of  
eq. (\ref{KPVMD}), we can get the corresponding linear slopes of $z_+$ and
$z_0$: $0.77\le -r_+\le 1.61$, $0.72\le -r_0 \le 1.17$ in the range 
of $0.1\le\eta_V\le 1.0$. These
values are not small if this kind of Taylor expansion is valid
here.  Unfortunately, it is not very easy to measure these quantities
experimentally.   On the other hand, we find that, $z_0$ is related to
$T_c^*$ through a linear relation eq. (\ref{Z0TC}).  Therefore, it is
expected that
a high-precision experimental analysis of the $T_c^*/m_K$ distribution
from DE contribution may be able to measure this energy dependence of the
amplitude.

\acknowledgments   

G.D. would like to thank the hospitality of the Center for Theoretical
Physics at MIT where part of this work was done, and the ''Bruno
Rossi" INFN-MIT exchange program.

\end{document}